# Understanding the Coupling Mechanism of Gold Nanostructures by Finite-Difference Time-Domain Method

Aditya K. Sahu* and Satyabrata Raj

*Department of Physical Sciences, Indian Institute of Science Education and Research Kolkata, Mohanpur, Nadia 741246, India.*
*Corresponding author
 E-mail: aks16rs023@iiserkol.ac.in

**Abstract**

Gold nanoparticle assemblies show a strong plasmonic response due to the combined effects of the individual nanoparticles' plasmon modes. Increasing the number of nanoparticles in structured assemblies leads to significant shifts in the optical and physical properties. We use Finite-Difference Time-Domain (FDTD) simulations to analyze the electromagnetic response of structurally ordered gold nanorods in monomer and dimer configurations. The plasmonic coupling between nanorods in monomers or dimers configurations provides a unique technique for tuning the spectrum intensity, spatial distribution, and polarisation of local electric fields within and surrounding nanostructures. Our study shows an exponential coupling behavior when two gold nanorods are assembled in end-to-end and side-by-side dimer configurations with a small separation distance. The maximum electric field in the gaps between adjacent nanorods in end-to-end dimer configuration describes a more significant enhancement factor than the individual gold nanorod. Our FDTD simulation on dimer in end-to-end assembly for small separation distance up to ~ 40 nm can well explain the observed experimental growth dynamics of gold nanorods.

## 1. Introduction

Over the last decade, significant progress has been made in developing new strategies for designing new plasmonic nano-assemblies with improved optical properties. Thereby it expands the scope of applications of such controlled nanostructure assemblies in sensing, imaging, nonlinear optics, catalysis, and diagnostics, etc. [1-3]. Gold nanoparticles have distinct optical properties in the visible and near-infrared parts of the electromagnetic spectrum, which can be helpful in various applications. Localized Surface Plasmon Resonance (LSPR) is a unique optical property of gold nanoparticles. LSPR arises due to the polarization of the free electrons within the nanoparticle, which resonant with the electric field of the light. The effect of shape and the separation among the nanoparticles are significant in the LSPR of gold nanoparticles [4]. In the nano-systems, the strong electromagnetic field on the nanoparticle surface arises due to the coherent collective excitation of the free electrons at the corresponding resonant wavelength of incident light [5].

In particular, because of their anisotropic form and plasmonic resonances, gold nanorods are significant in end-to-end assembly and other geometries to yield substantial field enhancements in the near-infrared region. The response of plasmonic nanoparticles to the LSPR depends on the particle's size, shape, composition, and the surrounding environment [6-10]. In the case of gold nanorods, it can be tuned from the visible to the near-infrared spectral region by varying the aspect ratio of the nanorods. The LSPR position can be modified for a given solar cell or Surface Enhanced Raman Spectroscopy (SERS) applications to improve their performance [11]. The reactivity of the LSPR peak position to the surrounding medium of nanoparticles has been used in Nanomedicine, sensors, and many applications [12]. More recent studies have shown that the near-field coupling of neighboring plasmonic nanoparticles can further improve the sensitivity to LSPR. When the separation distance between two nanoparticles is increased, a strong electromagnetic field is formed between the two nanoparticles, and a characteristic blue shift in the LSPR is noticed. As a result of the high electromagnetic field, SERS has been able to identify molecular species more effectively than single plasmonic nanoparticles [13]. In addition, the response of LSPR to nanoparticle interparticle separation distance is utilized as a scalable and optimised nanoscale molecular sensing and distance determination technique [14,15]. The plasmon oscillation induced by the light field leads to a strong electric field confined to the nanoparticle surface. Because of this induced plasmon oscillation, the near-field is strengthened on the nanoparticle surface relative to the incident field at their LSPR frequencies [16]. The near-field coupling of individual nanoparticle resonances in assembled proximal metal nanoparticles significantly impacts the LSPR. Such plasmon interaction among the nanoparticles provides a unique technique for tuning the spectrum, intensity, spatial distribution, and polarisation of local electric fields within and surrounding the nanostructures [17,18]. With this aim, several approaches have been exploited and reported in the literature over the past decades to design controlled assemblies of plasmonic nanostructures.

According to recent studies, the geometries of nanoparticles that exhibit high electromagnetic fields are exciting and potential candidates for prospective nano-sensors. Nonetheless, most current studies involving the pairing of the structure of nanoparticles were limited to separation distances greater than particle size. Hence the coupling behaviour of nanoparticles having various geometries such as a dimer, trimer, and so on with potential alignment as end-to-end and side-by-side with different separation gaps than particle size has yet to be explored in great detail. The present study explains the changes in the plasmonic spectrum and field maps for gold nanorods of different orientations based on their separation distances.

## 2. Methods

### 2.1 Synthesis and Characterization

**2.1.1 Materials:** Tetrachloroauric acid ($HAuCl_4 \cdot 3H_2O$) (99.9 %), Silver Nitrate ($AgNO_3$) (99.99 %), and ascorbic acid (99 %) were all purchased from Sigma-Aldrich. Cetyl trimethylammonium bromide (CTAB) (≥98%) and Sodium borohydride ($NaBH_4$) (99.99%) were purchased from Merck for the development of gold nanoparticles.

### 2.1.2 Synthesis of Gold Nanorods

As mentioned in the literature, the seed-mediated synthesis procedure was employed to synthesize gold nanorod solution [8-10]. All of the reactions were carried out in distilled water at room temperature. $NaBH_4$ (0.60 mL, 0.010M) was added to a solution mixture of CTAB (5 mL, 0.20M) and $HAuCl_4$ (5 mL, 0.0005M) while stirring continuously, resulting in a brownish-yellow solution called as seed solution. The mixture was stirred for few minutes and maintained at 25 ºC. CTAB (5 ml, 0.2M), $HAuCl_4$ (5 ml, 1mM), $AgNO_3$ (100 μl, 0.0064M), and ascorbic acid (85 μl, 0.0788M-2M) were mixed together to make a growth solution. The color of the solution turns colorless when ascorbic acid is added. The seed solution was added to the growth solution at 25 ºC to start the growth of gold nanorods, and the color of the solution gradually altered with time.

### 2.1.3 Characterization

The extinction spectra of colloidal solutions were measured using a UV-Vis-NIR HITACHI U-4100 spectrometer to determine the extinction spectrum of gold nanorods in solution and analyze the temporal evolution of the growth of gold nanostructures. Au nanorods' scanning electron microscope (SEM) image was produced using a Carl Zeiss Field Emission Scanning Electron Microscope. The different ensembles of nanoparticles can be seen from the scanning electron microscope images.

### 2.2 Simulation Methodology

The Finite-Difference Time-Domain (FDTD) approach was applied to model the optical characteristics and interactions between gold nanoparticles with different geometries [19]. FDTD is an effective and commonly used theoretical technology to model the optical properties of various shapes and sizes of plasmonic nanoparticles. To predict the optical properties of metallic nanoparticles, Maxwell equations, describing electromagnetic components of the electromagnetic wave with quasi-free charges in metallic nanoparticles, need to be resolved. For particles far smaller than the wavelength of a colloidal metal solution, Gustav Mie [20] was the first to propose a hypothesis that connects particle size and shape to optical characteristics. Switching to spherical coordinates solved Maxwell equations for spherical particles with lower dimensions than the wavelength, yielding an analytical solution for scattering and absorption coefficients of metal nanospheres in any medium.

FDTD simulations can simulate nanoparticles with any shape, and this method is particularly useful for non-spherical nanoparticles synthesized by chemical synthesis methods, such as nanorods, nanotubes, nanoprisms, and nanostars. Single nanoparticle optical characteristics, including absorption, scattering, and extinction cross-sections, are simulated. The nanoparticle electrical field enhancement is also visualized to correlate the observed optical properties with the experiment. A commercial program from Lumerical is used in the present work to investigate gold nanoparticle plasmon resonances [19]. We used small mesh cells to increase simulation accuracy. We have studied the nanoparticles in two steps (i) one nanoparticle is analyzed in various media, including water, and (ii) more nanoparticles with a specific regular pattern. For non-interacting nanoparticles, such as electromagnetically separated nanoparticles, the periodic pattern is adequate. The augmented electric fields will pair and affect the localized plasmon resonance when the spacing between nanoparticles is only a few nanometers. The primary goal of single nanoparticle analysis is to acquire absorption, scattering, and extinction spectra. We used the total field scattered field (TFSF) source and experimental Johnson and Christy data to simulate the complex permittivity of gold [21]. The absorption monitor should be placed around the particle within the TFSF source zone, while the scattering monitor should be placed outside. The incident fields are subtracted from the total fields outside the TFSF zone, leaving only the field scattered by the particle. The enhancement of the electric field, both surrounding and inside the particle, is observed after the absorption and scattering peak positions are measured for each form to understand the development and stability of the peaks.

## 3. Results and Discussion

It is interesting to study elongated rod-shaped nanoparticles because they exhibit two two plasmon resonance oscillation modes. Gold nanorods have plasmon oscillations along with the longitudinal and transverse directions. The transverse oscillation has a maximum of about 520 nm, comparable to the resonance in gold nanospheres, and blue-shifted compared to the longitudinal oscillation in the spectrum. The position of the longitudinal plasmon resonance can be tuned according to the oscillation strength, which is determined by the nanorod's aspect ratio [22]. The high oscillaton strength of the plasmons leads to very strong inter-particle plasmon coupling. Because the rod is more polarised in the long axis, the longitudinal mode is significantly stronger than the transverse mode [23]. When it comes to sensing applications, the dielectric sensibility of plasmonic nanoparticles can be improved by optimizing plasmon coupling. The interest arises on the rod-shaped particles as two rods can pair up in two different ways: either by favorably bonding (when rods are oriented end-to-end) or by anti-bonding (when rods are placed side-by-side) [24]. Other assemblies of nanorods are possible and demonstrated schematically in Fig. 1a. Our interest lies mainly in strongly coupled bonding mode. The light polarization direction was maintained along the interparticle axis to explore the interparticle bonding of the long axis modes as a function of particle spacing.

The solution-phase assembly of gold nanoparticles provides a basic qualitative explanation of plasmonic coupling. Gold nanorods display rich colors due to the almost linear dependence of the longitudinal plasmon resonance wavelength, which highly depends upon the length-to-diameter aspect ratio. The colors of nanorods change for different aspect ratios, surrounding medium, and coupling among nanorods, etc. It has been found that the LSPR maximum can be tuned by changing only the ascorbic acid concentration and keeping all the synthesis parameters identical. In our experiment, the ascorbic acid acts as a reducing agent and a capping agent to help control the anisotropic growth of gold nanorods. This is indeed the scenario at ascorbic acid concentration 0.1 M where dog-bone structures were formed. Fig. 1b represents the absorption spectra of solution showing LSPR peak position around 675 nm, 743 nm, 683 nm, and 703 nm for the ascorbic acid concentration of 0.0788 M, 0.1M, 0.15M, and 0.2M, respectively. Fig. 1c shows the side-by-side assembly of nanorods in the SEM images of the gold nanorods for different ascorbic acid concentrations. Many assembled nanorods structures can be modeled as mentioned before (shown in Fig. 1a). To understand its coupling effect, we have extensively studied monomer and dimer with end to end and side-by-side geometry.

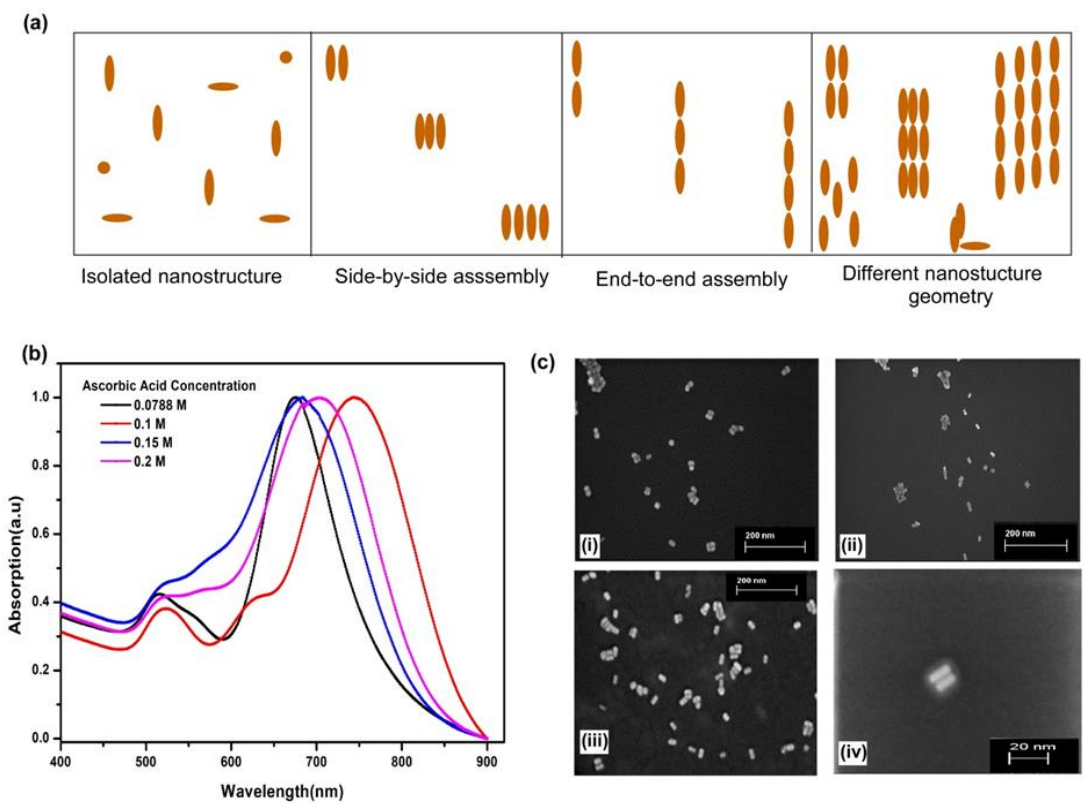

**Fig. 1.** (a) Schematic presentation of different geometry of nanostructures. (b) The effect of ascorbic acid concentration on the absorption spectrum of gold nanorods obtained from UV-Vis spectrometer. (c) SEM images of the gold nanorods with varying ascorbic acid concentrations showing different structures like side-by-side assembly.

The different configurations of the nanorods provide excellent tunability for resonance, and it has been found that the plasmon resonance is red-shifted by increasing the length of the nanorods. We have analyzed the properties of gold nanorods with a fixed width of 30 nm and a range of lengths from 40 ~ 150 nm. The aspect ratio-dependent absorption cross-section of the longitudinal plasmon resonance of the gold nanorods due to the light polarized along the long axis of the nanorod is shown in Fig. 2a. We observed that the plasmon resonance strength increases and red-shifted with the increase in the aspect ratio of nanorods. For the length of ~ 150 nm, the resonance can be tuned to the near-infrared region for the gold nanorods. The numerical results show that the extinction cross-section of a gold nanorod is sensitive to the rod's aspect ratio and sensitive to their orientation in the assembly. Electric field enhancement occurs mainly at the end cap of gold nanorods due to the LSPR, as explained later (shown in Fig. 3a). It is noteworthy to mention that the irradiating light polarization is parallel to the long axis of the nanorods. When gold nanoparticles are placed close together, the localized surface plasmon resonances in individual particles pair with one another, resulting in a more profound plasmon resonance than when the particles are isolated (see Fig. 2b). The field analysis shows a high factor (~ 50 for 150 nm rods for single nanorods, whereas ~ 1 order more than single nanorods for dimer separated by 1 nm) of enhancement at the ends of nanorods and the field increases exponentially with the increase in the aspect ratio of the nanorods (shown in Fig. 2c). Fig. 2d shows the variation of LSPR peak position with the aspect ratio of nanorods. It has been found that the LSPR peak redshifts in dimer assembly compared to single nanorods. When the particle aspect ratio is increased, the intra-particle coulomb restoring force decreases, single-particle polarizability (and dipole moment) increases, and the near-field interaction becomes more significant. As a result, the fundamental plasmon wavelength shift is more substantial with greater aspect ratio dimer formation.

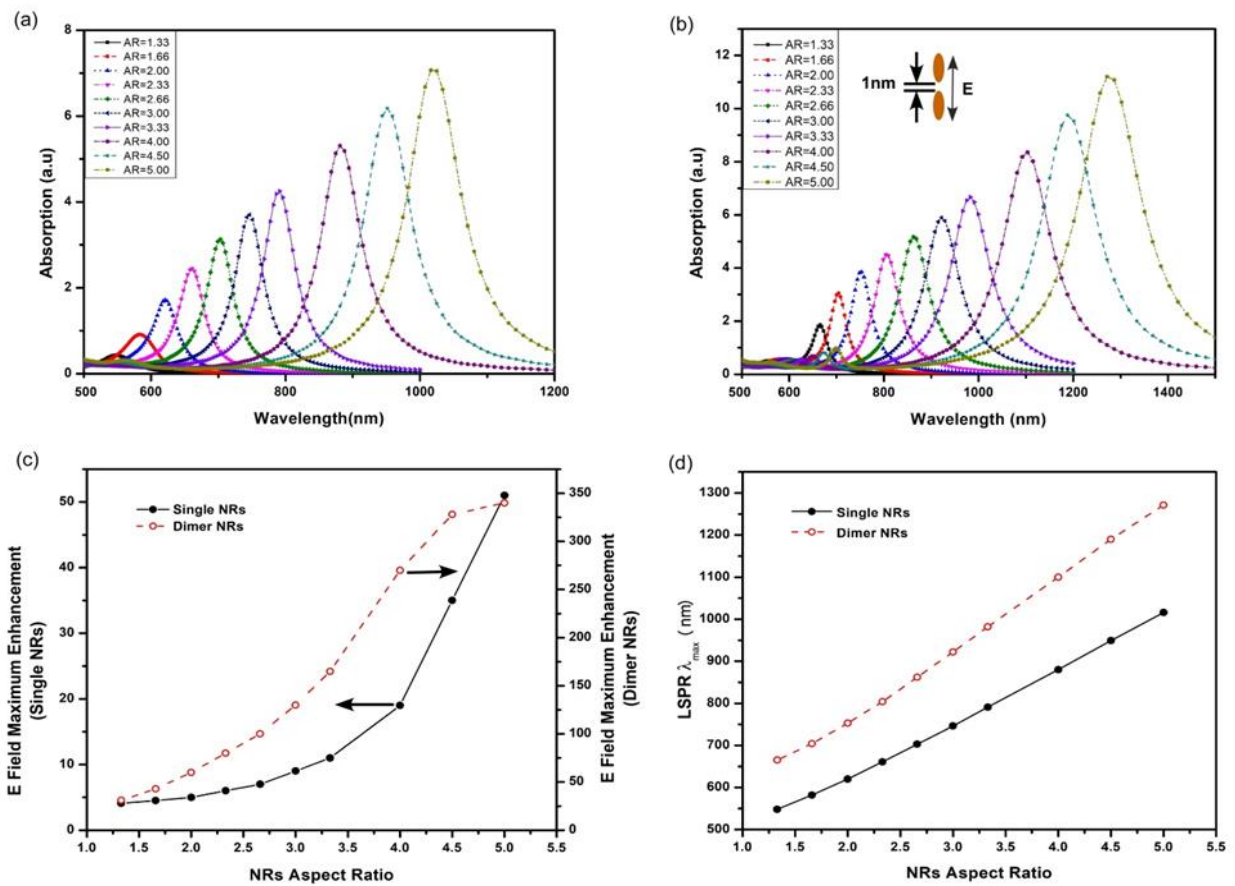

**Fig. 2.** Absorption cross-sections vs. aspect ratio of (a) single nanorods (b) nanorods in dimer structure. (c) Variation of the intensity of maximum electric field enhancement of single and dimer nanorods. (d) Comparison of LSPR position in single and dimer structure w.r.t aspect ratio of nanorods.

Fig. 3a displays the aspect ratio dependent electric field mapping of an isolated single nanorod. A significant enhancement factor has been recorded at both ends of a single nanorod, which increases exponentially with increasing aspect ratio. Fig. 3b displays the aspect ratio dependence electric field mapping for comparable nanorods producing dimer structures. The field profiles clearly show that the regions with the higher electric field near the connectors are known as hot spots. As previously stated (shown in Fig. 2c), the field enhancement factor increases as the aspect ratio increases. Because nanorods are anisotropic, charges appear to concentrate in the rounded corners of the long ends, resulting in significant field increases in those places [25]. Under longitudinal polarisation, the dipolar resonance formed between rods rises with the growth of the aspect ratio. As a result, the rod with the largest aspect ratio produces the most remarkable electric field enhancement. As well as being affected by the size, the enhancement factor also depends on how well two equivalent nanorods can couple together, which is controlled by the particle spacing.

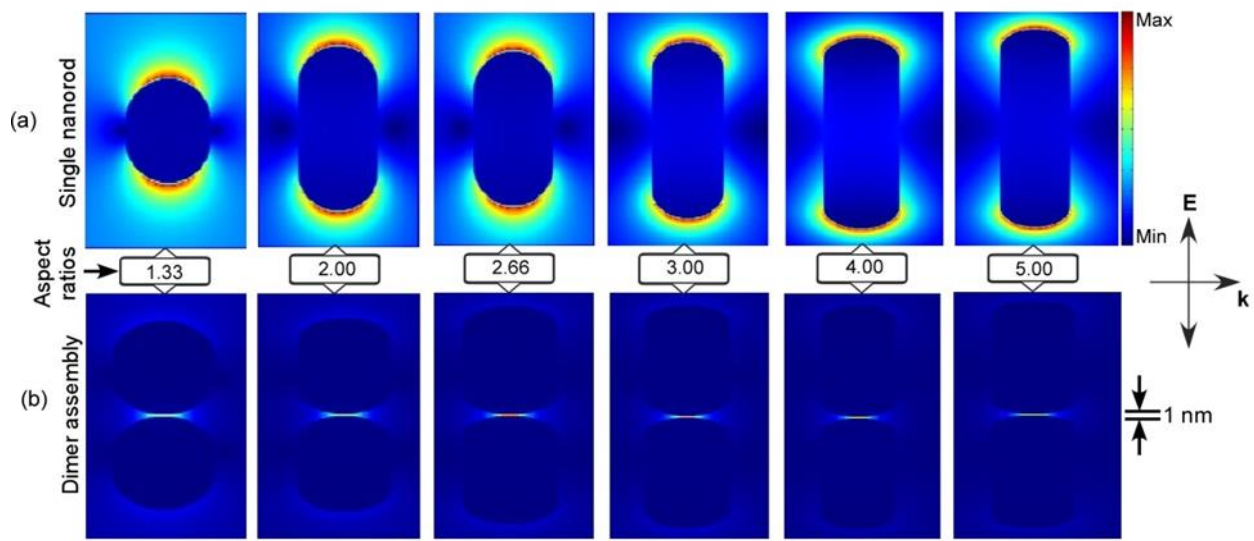

**Fig. 3.** Electric field intensity distribution mapping with different aspect ratios for (a) single gold nanorod and (b) dimer nanorods assembly placed end-to-end with a 1 nm inter-particle separation gap.

Now, we explore the effect of the separation distance of nanorods having a constant aspect ratio on plasmon resonance in a dimer assembly. As explained before, the nanorods can be constructed in one of two ways: end-to-end or side-by-side. The effect of inter-particle distance on plasmon coupling is obtained from FDTD simulation of gold nanoparticles with symmetrically varying inter-particle separation. As demonstrated in Fig. 4a, the plasmon resonance wavelength falls exponentially as the inter-particle distance, $d$, increases. The magnitude of the plasmon resonance significantly decreases as the interparticle distance increases, indicating that the strength of the interaction between nanorods in dimer decreases exponentially over the interparticle range. Two prominent plasmonic modes (ex. as marked $R_1$ and $R_2$ for gap ~ 1 nm) are visible in Fig. 4a and significantly depend upon the separation gap of nanorods connected end-to-end in dimer assembly. As the separation distance increases, peak $R_1$ strength decreases drastically and disappears as the separation distances increase more than 3 nm. Though the intensity of peak $R_2$ decreases with increases d from 1 - 40 nm, the rate of intensity fall for $R_1$ and $R_2$ is quite different for small d value. It seems that two plasmonic coupling domains result in the enhancement of two different plasmonic modes. For d value greater than 3 nm, the significant plasmonic resonance ($R_2$) is caused by the near-field coupling between two individual gold nanorods. For a small d value, $R_1$ becomes the plasmonic resonance due to the effect of the separation gap.

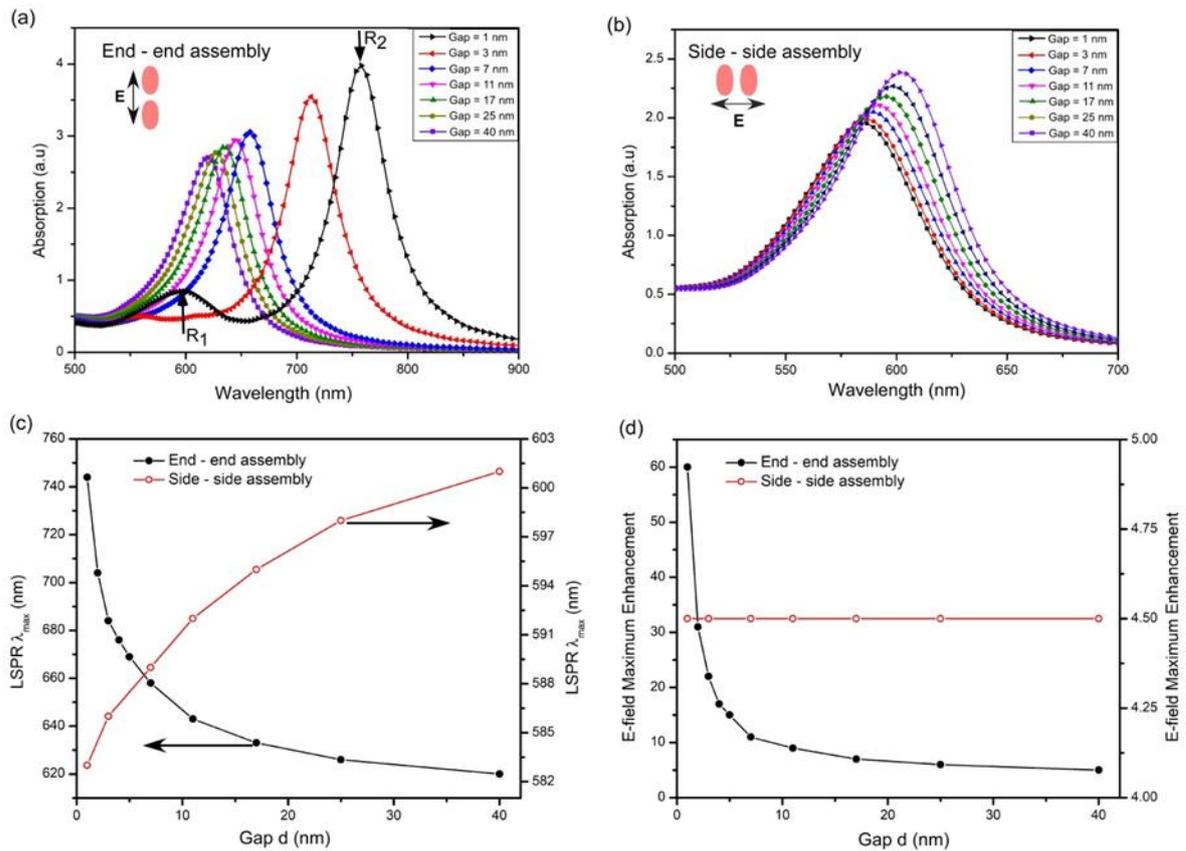

**Fig. 4.** Absorption spectrum of gold nanorods (aspect ratio ~ 1.81) w.r.t separation gap, d in (a) end-to-end and (b) side-by-side dimer assembly. Variation of plasmon resonance (c) peak position and (d) intensity w.r.t nanorods gap, d in different dimer assembly configurations.

Interestingly for side-by-side orientation, the resonance peak shifts towards a higher wavelength as the gap, d increases from 1 to 40 nm, as shown in Fig. 4b. The trend is exactly the opposite (shown in Fig. 4c) to the end-to-end orientation of nanorods. The inter-particle interaction is highly attractive in the end-to-end configuration, leading to a significant blue shift. On the other hand, the side-by-side arrangement has a small redshift due to a minor repulsive attraction between the electronic dipoles in the dimer assembly. We also observe an opposite behavior (shown in Fig. 4d) in electric field enhancement (which is related to the resonance peak intensity of the plasmon band) of different dimer configurations due to the diffraction coupling in the nanorods pairs when the gap distance increases [26,27].

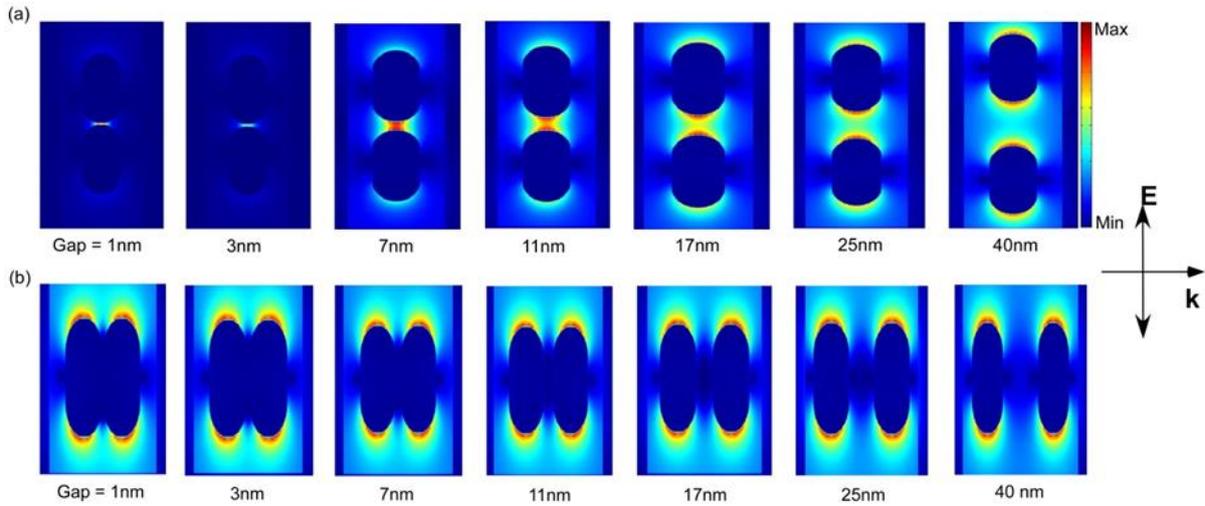

**Fig. 5.** The electric field intensity distribution with different interparticle spacing, d of dimer nanorods with an aspect ratio of ~ 1.81 for (a) end-to-end assembly and (b) side-by-side assembly.

Fig. 5 shows the interparticle spacing d dependent field profile of dimer nanorods with an aspect ratio of ~ 1.81. The gap, d between the nanorods is varied from 1 – 40 nm. When d increases, the electric field around the two rods decreases, demonstrating that the resonance of the two rods decreases due to the mutual interference-effect. Fig. 5a shows a high field enhancement of ~ 60 for 1 nm close-spaced nanorods, whereas the enhancement factor decreases exponentially to ~ 5 for the 40 nm gap size. The enhancements in the field become appreciable when nanorods are closer than 40 nm; this allows the local field to be coupled and leads to higher field levels in the region between the nanorods. If nanorods are placed side by side assembly, the field patterns at their $\lambda_{max}$ show invariant dependency on the interparticle spacing, d. The dimer acts as two isolated monomers when the incident light is polarized perpendicularly (shown in Fig. 5b) due to the reduced plasmon coupling between gold nanorods. We can observe that the end-to-end coupling is significantly stronger than side-by-side coupling, and therefore the strength of the electric field is much higher at the end-to-end coupling.

It has been experimentally observed that during the growth of gold nanorods, the LSPR peak first red-shifted and then switched towards the blue shift, as shown in Figs. 6a and b. Hence we carried out theoretical calculations using the FDTD simulation approach to understand better the plasmonic properties of the time-dependent gold nanorod growth dynamics as observed experimentally. As the nanorods grow from their initial stage of spheres in solution, it was observed that as the seed-solution is added to the growth solution, the solution undergoes a color change with time progress. The growth solution is initially colorless. As time goes on, it changes to wine red, then violet, and finally settles in various colors like violet, dark blue, cyan, or ruby red. For the study on the system's evolution with time, time-dependent absorption spectrums were taken at regular intervals. Immediately after adding the seed solution to the growth solution, the resulting solution was transferred to a quartz cuvette and loaded into a UV-Vis NIR spectrophotometer. The absorption spectrum was recorded at regular intervals of 4 ~ 70 minutes, depending upon the reaction thermodynamics. The initial 4 minutes time lag was due to the pre-measurement sample loading process into the spectrometer. Two peaks in the UV-Vis spectra evolve as the time of growth reaction increases. The transverse surface plasmon resonance (TSPR) peak is observed around ~ 520 nm, whereas the position and intensity of the LSPR peak vary with the progress of the growth reaction. The overall behavior of the particles leads to the LSPR peak position starting at around ~ 728 nm and increasing to ~ 745 nm, then decreasing and finally stabilizing at about ~ 634 nm, as shown in Fig. 6a and b. This behavior contradicts the fundamental understanding of the mechanism of growth. The rod is believed to grow in solutions from spheres, and the LSPR peak should display a red-shift, provided other parameters are unchanged. The red-shift nature is that the electronic confinement of the long axis of the rod decreases as the length of the nanorod increases. So the maximum longitudinal peak should shift towards the higher wavelengths. However, we observe that the LSPR peak undergoes a red-shift only within the initial ~ 10 minutes of growth reaction, followed by a blue-shift after that, and ultimately saturates after ~ 50 minutes. From the FESEM images, we can see several particles with variable particle separation domains, such as widely separated, closely separated, touching, and non-touching particles, and overlapped particles, each of which has its own set of characteristics. They behave independently when the interparticle separation is large. As particle separation decreases, the local field to be coupled and leads to higher field enhancement in the region between the nanorods. The near fields in the gaps parts are enhanced, as previously mentioned in Fig. 5. We show theoretically that when particles are separated but close together, a new regime emerges. As a result, we extend a better understanding of these dimer regimes of nearly touching and separated dimers, which is significant for interpreting recent developments on the growth dynamics of gold nanorods.

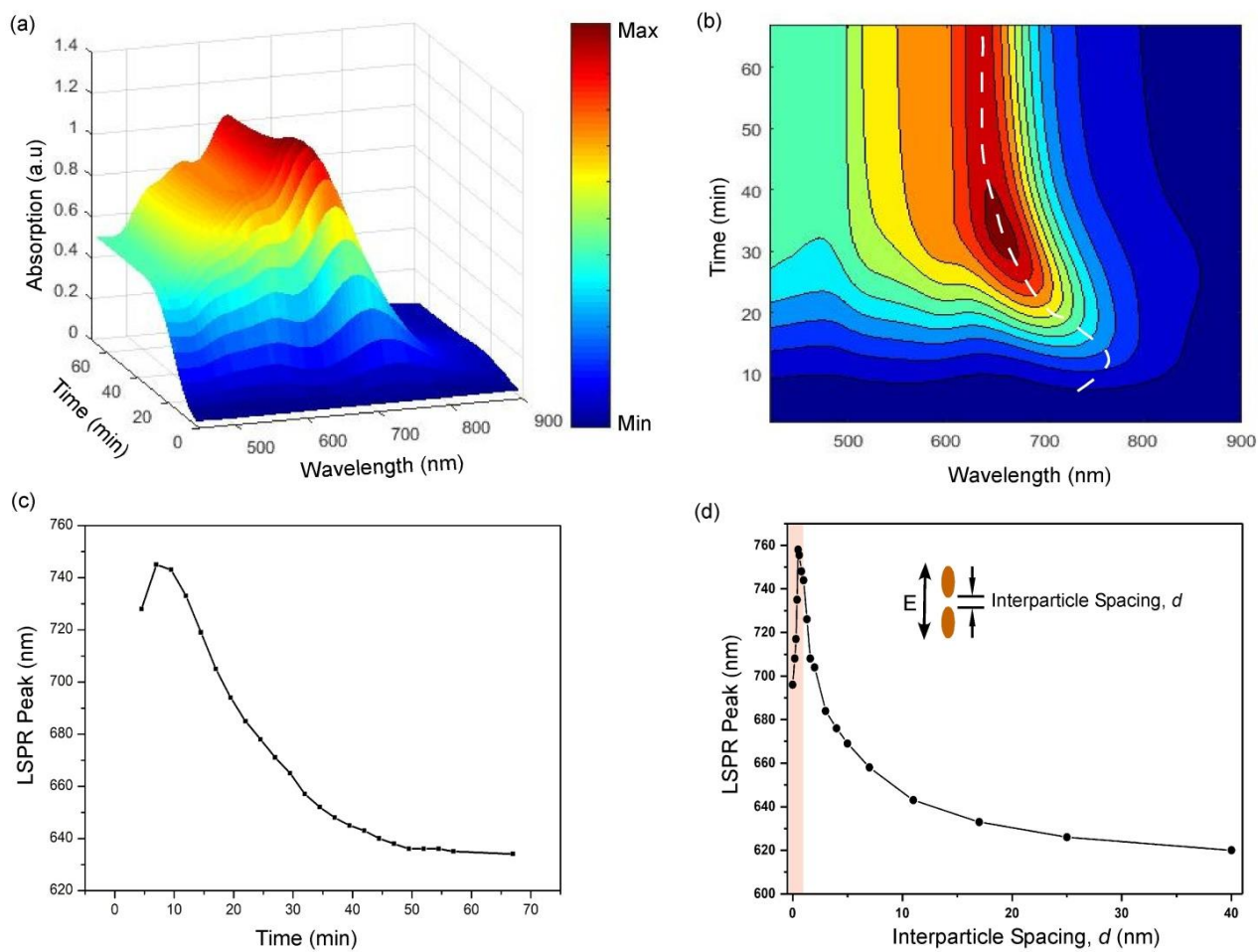

**Fig. 6**. (a) Time-dependent absorption spectra of gold nanorod while growth is in progress (b) Top view of the variation of LSPR peak position as a function of growth reaction time (white dashed line directs an observer for development of the peak position). (c) Change in the LSPR peak position in time extrapolated from Fig. 6b. (d) LSPR peak position vs. interparticle separation gap in the end-to-end assembly of nanorods for aspect ratio 1.81 simulated by the FDTD method. The shaded color region shows the result for d from 0 to 0.5 nm, an unrealistic value as the Au-Au bond length varies from 0.28 - 0.32 nm [28].

Fig. 6c demonstrates the experimental change of LSPR peak position during the growth of nanorods. In contrast, Fig. 6d provides the change in LSPR peak position with the interparticle spacing of gold nanorods with an aspect ratio of ~ 1.81, forming dimer configurations. One can observe the similarity between the experimental data (shown in Fig. 6c) with the simulated one (shown in Fig. 6d). We have calculated the absorption spectrum by changing the gap from 0 to 40 nm in dimer structure with nanorods in the end-to-end assembly in the surrounding environment of the refractive index around 1.33. We note that the plasmon resonance peak moves towards a higher wavelength (i.e., red-shifts) by increasing the interparticle distance, d from 0 to 0.5 nm, as shown in the shaded color region. Later peak position moves towards a lower wavelength till 620 nm (i.e., blue-shifts) by further increasing d from 0.5 nm to 40 nm. The near-field of each nanoparticle can interact with each other when they are brought close enough. The electric field near each particle exhibits a dipole-like structure at large separations, with intense fields enhancement in the orientations of the interparticle axis. The electric field enhancement in the interparticle area increases as the distance between particles decreases due to the coupling of local fields. As the field coupling across the interparticle separation region becomes significant, the dipole model disappears at short enough separations because the end-to-end interactions around the particles cannot be operated. The interaction between charge accumulates when particles collide forces the excitation spectra to evolve after contact and the modes to redshift due to mode decoupling. Touching dimers operate as a single elongated particle, and the dimer's modes behave similarly to the modes of a nanorod. It is important to note that there is a clear transition between the modes before and after contact. Experimentally, the same shift can be seen. The plasmonic mode appears to evolve at the transition point after the contacting particles are separated and act as independent charged modes. Contrary to field enhancement at a point contact, when particles with flat ends touch or separate, i.e., side-by-side ensembles, the near field coupling at the junction is neutralized, showing a constant field enhancement of ~ 4.5 (as discussed in Fig. 4d). The lowest physical dipole mode of the particles evolves as plasmonic characteristics of the separated dimer in a side-by-side ensemble (~ 580-600 nm range as mentioned in Fig. 4c). As a result, the contribution of side-by-side ensemble dimeric nanoparticles is not considered to understand our experimental results.

The plasmon oscillations of nanoparticles can be coupled due to this near-field interaction. This plasmon coupling modulates the correlated nanoparticle system's LSPR. The plasmon resonance band shows a continuous blue shift till 620 nm for a 40 nm gap because of plasmonic interaction among particles based on a dipole-dipole coupling. The red-shift in the resonance spectrum for 0 - 0.5 nm distances indicates that the two gold nanorod ends are bound to each other. FDTD simulation predicts that these smaller nano-gap structures cause more substantial plasmonic near-field enhancement in the nano-gap position. The redshift shows a favorable coupling of the nanoparticles' proximal plasmonic oscillation, implying that less energy is needed to induce the oscillation modes coupling the plasmon. These results are consistent with the field map of the induced electric fields in touching, nearly-touching, and well-separated nanorods. The 0.5 nm gap shows the nearly-touching of the ends acts as an interface between the particle properties of the touching and well-separated nanorods. Through numerical simulation, our approach provides some insight into the almost touching regime, resulting in an explanation of the evolution of the plasmonic modes and induced electric field enhancement in gold nanorods, but the simulated distance, d from 0 to 0.5 nm when the plasmon resonance peak moves towards a higher wavelength, is unrealistic values. The bond length of Au-Au in the Au cluster varies from 0.28 - 0.32 nm, so the separation distance below 0.5 is not realistic.[28] The computed response explains the recent experimental behavior of the growth dynamics of gold nanorods and can be used to construct sensors based on gold nanoparticles and plasmonic electronic components.

## 4. Conclusion

We have investigated the gold nanorod LSPR and field enhancement factor by using the FDTD method in this study. Tuning plasmon resonance using various geometries such as single gold nanorods and gold nanorods in dimer configuration has been studied extensively. We found plasmonic material offers better field enhancement based on the nanostructure's shape, size, and configuration. The enhancement factor dramatically increases in the nanorods pairing and depends on axes between similar nanorod forming dimers. The end-to-end dimer configuration shows more enhancements compared to the side-by-side configuration. We demonstrated a significant blue shift in the plasmon resonance coupling between two gold nanorods with interparticle separation of 1 ~ 40 nm when kept in end-to-end assembly compared to a side-by-side orientation. A blue shift for separation distance (between 0.5 ~ 40 nm) explains the experimental growth dynamics of nanorods. These plasmonic coupling of nanorods of various assembly (i.e., monomers and dimers, etc.) provides a unique technique for tuning the spectrum, intensity, spatial distribution, and polarization of local electric fields within and surrounding nanostructures.

**Funding**

The author, AKS, receive financial support from CSIR, the Government of India.

**CRediT authorship contribution statement**

**Aditya K. Sahu**: Conceptualization, Methodology, Experiment, Software, Data analysis, Investigation, Writing - original draft, Writing, Writing - reviewing & editing. **Satyabrata Raj**: Resources, Writing - reviewing & editing, Funding acquisition, Supervision.

**Declaration of competing interests**

The authors declare that they have no known competing financial interests or personal relationships that could have appeared to influence the work reported in this paper.

**Availability of Data and Material**

The data that support the findings of this study are available from the corresponding author upon reasonable request.